\documentclass[11pt]{article}
\usepackage{blois,epsfig}

\def\gtwid{\mathrel{\raise.3ex\hbox{$>$\kern-.75em\lower1ex\hbox{$\sim$}}}}
\def\ltwid{\mathrel{\raise.3ex\hbox{$<$\kern-.75em\lower1ex\hbox{$\sim$}}}}
\def\\{\hfil\break}

\def\apj{{\em ApJ} }
\def\apjl{{\em ApJ Lett.} }

\def\be{\begin{equation}}
\def\ee{\end{equation}}
\def\bea{\begin{eqnarray}}
\def\eea{\end{eqnarray}}

\begin{document}
\vspace*{4cm}
\title{Optimal Moments for Velocity Fields Analysis}

\author{ Hume A. Feldman$^\dagger$, Richard Watkins$^\ddagger$, Adrian Melott$^\dagger$ \& Will Chambers$^\dagger$}
\address{$^\dagger$Department of Physics \& Astronomy,
University of Kansas, Lawrence, KS 66045, USA\\
$^\ddagger$Department of Physics, Willamette University,
Salem, OR 97301, USA}

\maketitle\abstracts{
We describe a new method of overcoming
problems inherent in peculiar velocity surveys by using data compression as a filter with which to separate large--scale, linear flows from small--scale noise that biases the
results systematically.  We demonstrate the effectiveness of our method using realistic
catalogs of galaxy velocities drawn from N--body simulations.  Our tests
show that a likelihood analysis of simulated catalogs that uses all of
the information contained in the peculiar velocities results in a bias
in the estimation of the power spectrum shape parameter $\Gamma$ and
amplitude $\beta$, and that our method of analysis effectively removes
this bias.  We expect that this new method will cause peculiar velocity
surveys to re--emerge as a useful tool to determine cosmological
parameters.
}

%\section{Introduction}
%\label{sec-intro}

We introduce a
new method for the analysis of peculiar velocity surveys~\cite{PI,PII} that is a
significant improvement over previous methods.  In particular, our
formalism allows us to separate information about large--scale flows
from information about small scales, the latter can then be
discarded in the analysis.  By applying specific criteria, we are able
to retain the maximum information about large scales needed to place the
strongest constraints, while removing the bias that small--scale
information can introduce into the results.

%\section{The Formalism}
%\label{sec-formalism}

To analyze the observed line--of--sight velocities we assume that
N objects with positions $r_i$ and observed line--of--sight velocities
$v_i$ can be modeled as
\begin{equation}
v_i=\vec v(\vec r_i)\cdot \hat r_i+\delta _i
\end{equation}
where $ v(\vec r_i)$ is the linear velocity field and $\delta_i$ is the
noise which also accounts for the deviations from linear theory.  Assume the
noise is Gaussian with variance $\sigma_i^2+\sigma_*^2$ where $\sigma_i$
is the observational error and $\sigma_*$ is the contribution from
nonlinearity and other things we neglected (see ~\cite{fw94} for detail
analysis).  The covariance matrix can be written as
\begin{equation}
R_{ij}=\left\langle {v_i\,v_j} \right\rangle =R_{ij}^{(v)}+\delta _{ij}\left( {\sigma 
_i^2+\sigma _*^2} \right)
\qquad{\rm where}\qquad 
R_{ij}^{(v)}=\left\langle {\vec v(\vec r_i)\cdot \hat r_i\,\,\,\vec
v(\vec r_j)\cdot \hat r_j\,} \right\rangle\,.
\end{equation}

In linear theory we can express the velocity power spectrum in terms of
the density power spectrum and thus rewrite the above as
\begin{equation}
R_{ij}^{(v)}={{H^2f^2\left( {\Omega _0} \right)}\over {2\pi ^2}}\int {P(k)W_{ij}^2(k)dk}\,.
\end{equation}
The covariance matrix is a convolution of the density power spectrum and the squared 
tensor window function.

The probability distribution for the line--of--sight peculiar velocities is
\begin{equation}
L\left( {v_1,\cdots ,v_N;P(k)} \right)=\sqrt {\left| {R^{-1}} \right|}\exp \left( {{{-
v_iR_{ij}^{-1}v_j} \over 2}} \right)
\end{equation}
Alternately, given a set of velocities $ \left( {v_1,\cdots
,v_N}\right)$ we can have $ L\left( {v_1,\cdots ,v_N;P(k)}\right)$ to denote the
likelihood functional for the power spectrum.  Given a power spectrum
parameterized by some vector $\Theta = \left( {\theta_1,\cdots ,\theta_s}
\right)$ then $L\left( {v_1,\cdots ,v_N;\Theta}\right)$ is the
likelihood functional for the parameter $\Theta.$ The value of the
parameter vector that maximizes the likelihood we call $\Theta_{ML}$.  

Given a set of true parameter $\Theta_0$, we want a maximum likelihood
estimator $\langle\Theta_{ML}\rangle=\Theta_o$ then $\Theta_{ML}$ will
vary over different realizations of $\left( {v_1,\cdots ,v_N} \right)$.
We may characterized our parameters with the Mean
\(\langle(\theta_{ML})_i\rangle\)
and the variance
\(\Delta(\theta_{ML})^2_i=\langle(\theta_{ML})^2_i\rangle - 
\langle(\theta_{ML})_i\rangle^2\)
In the limit of large N:
\(\langle\Theta_{ML}\rangle=(\Theta_o)_i\)
and the variances are minimal.The variances for an unbiased estimators are: 
\begin{equation}	
\Delta \left( {\theta _{ML}} \right)_i\ge \left( {F_{ii}} \right)^{-1/2}
\end{equation}
which is the Cram\'er--Rao inequality. In the limit of large N
this becomes an equality, here we assume that this limit is satisfied. $F_{ii}$ is the trace of the Fisher matrix
\begin{equation}
F_{ij}=\left. {\left\langle {{{\partial ^2\left( {-\ln L} \right)} \over {\partial \theta 
	_i\partial \theta _j}}} \right\rangle } \right|_{\Theta =\Theta _0}
\end{equation}

If the velocities are Gaussianly distributed then the maximum likelihood
estimator $\Theta_{ML}$ is unbiased.  However, actual peculiar velocities
contain non--Gaussian contributions,  nonlinear contributions will lead to
$\Theta_{ML}$ being biased in an unpredictable way.  In order to recover
an unbiased estimator we utilize data compression methods.  We use these
methods to filter out unwanted information.

%\section{Results}
%\label{results}

The main purpose of the formalism we presented here and was
to allow the removal or filtering of small--scale noise while keeping the
large--scale signal. To test the success of the formalism we have created
synthetic surveys from simulations with known parameters, specifically,
$\Gamma$, the CDM power spectrum shape parameter, and $\beta$, its
amplitude. To compare our method with the usual maximum likelihood analysis method, we
reemphasize that the optimal moment analysis presented here allows for
two semi--independent methods of cleaning up a survey: a) Ordering the
moments by their eigenvalues and removing those with the largest eigenvalues b) Removing the
noisiest moments. In Fig.~1 we show the comparison between
choosing the modes least susceptible to small--scale signal; those that are least susceptible 
to small--scale signal {\bf and} are not noisy; and the full analysis (that is,
estimating the parameters using all the information). We see that
the full analysis fails to recover the ``true'' parameters by a
significant amount ($\approx4\sigma$ for no errors and $>2\sigma$ for
10\% errors). In contrast, the mode analysis recovers the values of the
parameters very well, with or without the removal of the noisy
moments.

In Fig.~2 we show the value of the estimated parameters as a
function of the $\Sigma\lambda^2$ where $\lambda$ is the eigenvalue, we see
that as the number of modes is increased, we get closer and closer to
the ``true'' value. When we keep more than the number of moments that
corresponds to the fulfillment of our criterion (solid vertical lines),
the values start diverging systematically from the ``true'' results.  This is due to
the fact that small--scale modes that have become nonlinear are
introducing a systematic bias.  This tendency of the full analysis to
systematically overestimate the parameter values can be seen for all values of the parameters.

\begin{figure}[ht]
\begin{center}
\psfig{figure=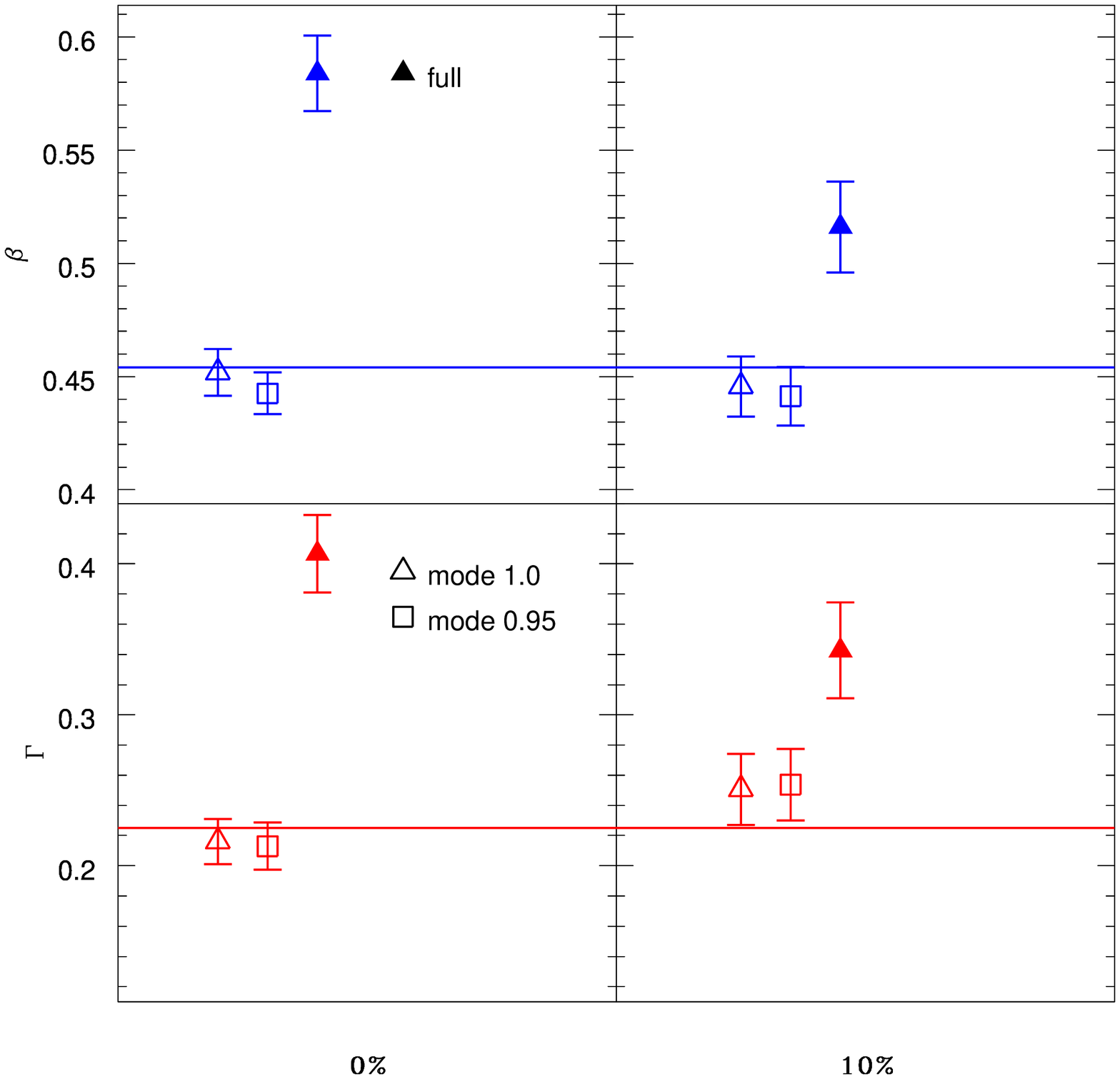,height=2.5in}\qquad\qquad\qquad
\psfig{figure=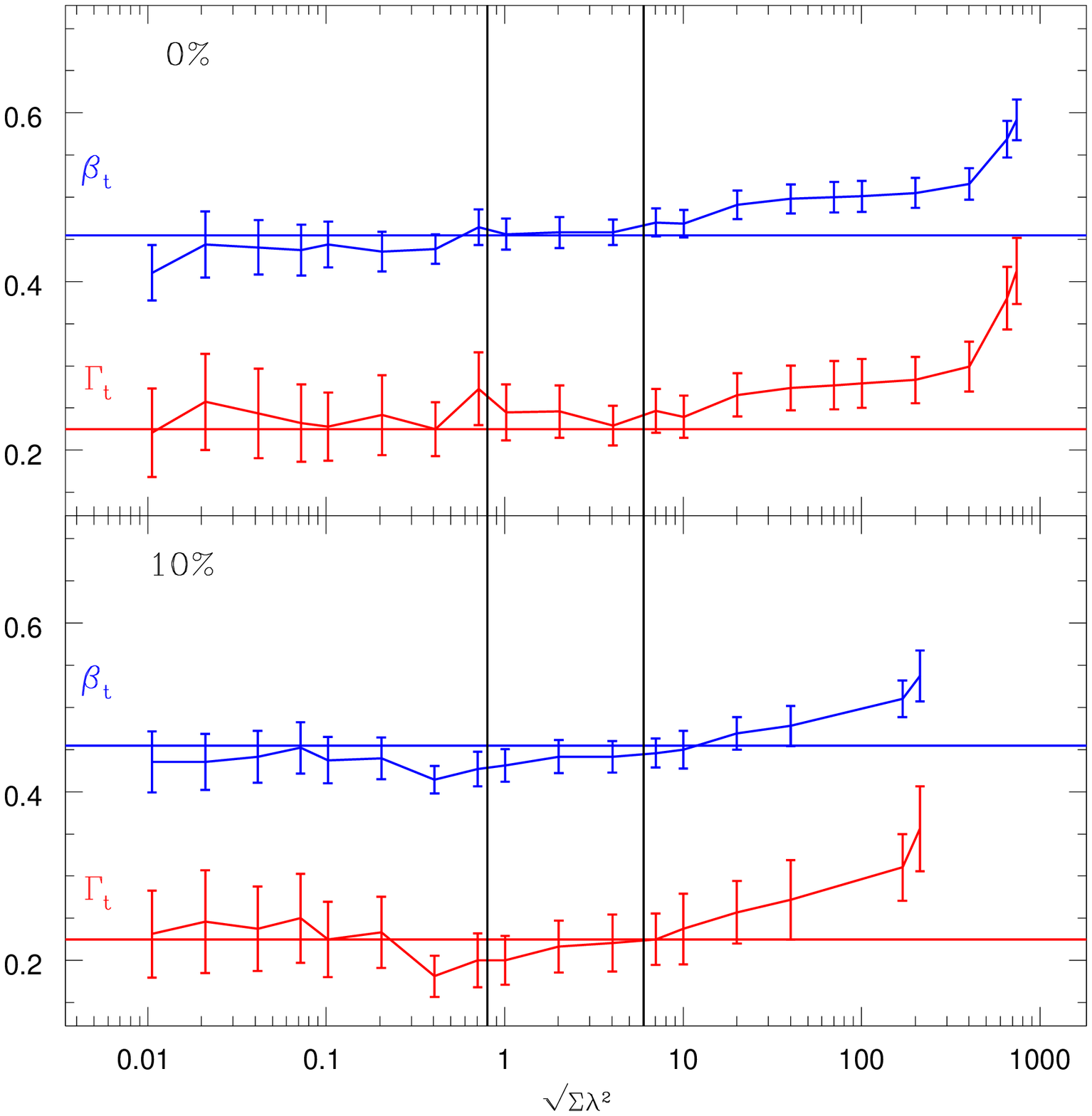,height=2.5in}
\end{center}
\caption{(left panel) A comparison between the mode analysis presented in this
paper and the usual full analysis.  In
the top two panels are the mean values and standard deviations of the
mean of $\beta$, the amplitude of the power spectrum. The bottom panels
are the results for estimating $\Gamma$, the shape parameter. In the
left panels we have the results for the analysis for a survey with no
errors whereas the right panels show the results for 10\% errors. The
solid symbols are the full analysis results and the empty ones are the
mode analysis. The triangles are the results without removing the noisy
moments, the rectangles are those where we removed the noisiest
moments. The horizontal lines are the ``true'' values of the parameters. \hfill\break
Figure 2: (right panel) The mean value of the estimated parameters from 81 catalogs extracted from the simulations as a function of the number of modes we keep. The top panel shows results for survey with no errors, the bottom panel shows the results with distance errors of
10\%. As the number of modes kept increases beyond the
criteria set, the estimators become systematically biased. The horizontal
lines are the ``true'' values of the parameters.
}
\label{fig-1-2}
\end{figure}

As was discussed in the text, the reason for the full analysis failure
to recover the ``true'' parameters when the mode analysis succeeds so
well can be shown by looking at the window functions themselves. In
Fig.~3 we show the normalized window functions $W_{n}(k)$
in arbitrary units vs. $k$, the wave number corresponding to the five
lowest eigenvalues and lowest noise (lower left panel). As we move up the panels we see the window
functions with larger noise components not removed, whereas when we
move to the right we see window functions corresponding to larger
eigenvalues. Here the reasons for the particular choices for our
criteria become clear. As
the eigenvalues or the noise level become large, the window functions
generally probe more small--scale and less of large--scale
modes. Since we are primarily interested in large--scale information,
discarding the noisy, high $\lambda$ modes allows us to remove
small--scale signal that might, and generally does, interfere with with our analysis.

\setcounter{figure}{2}
\begin{figure}[ht] 
\begin{center}
\psfig{figure=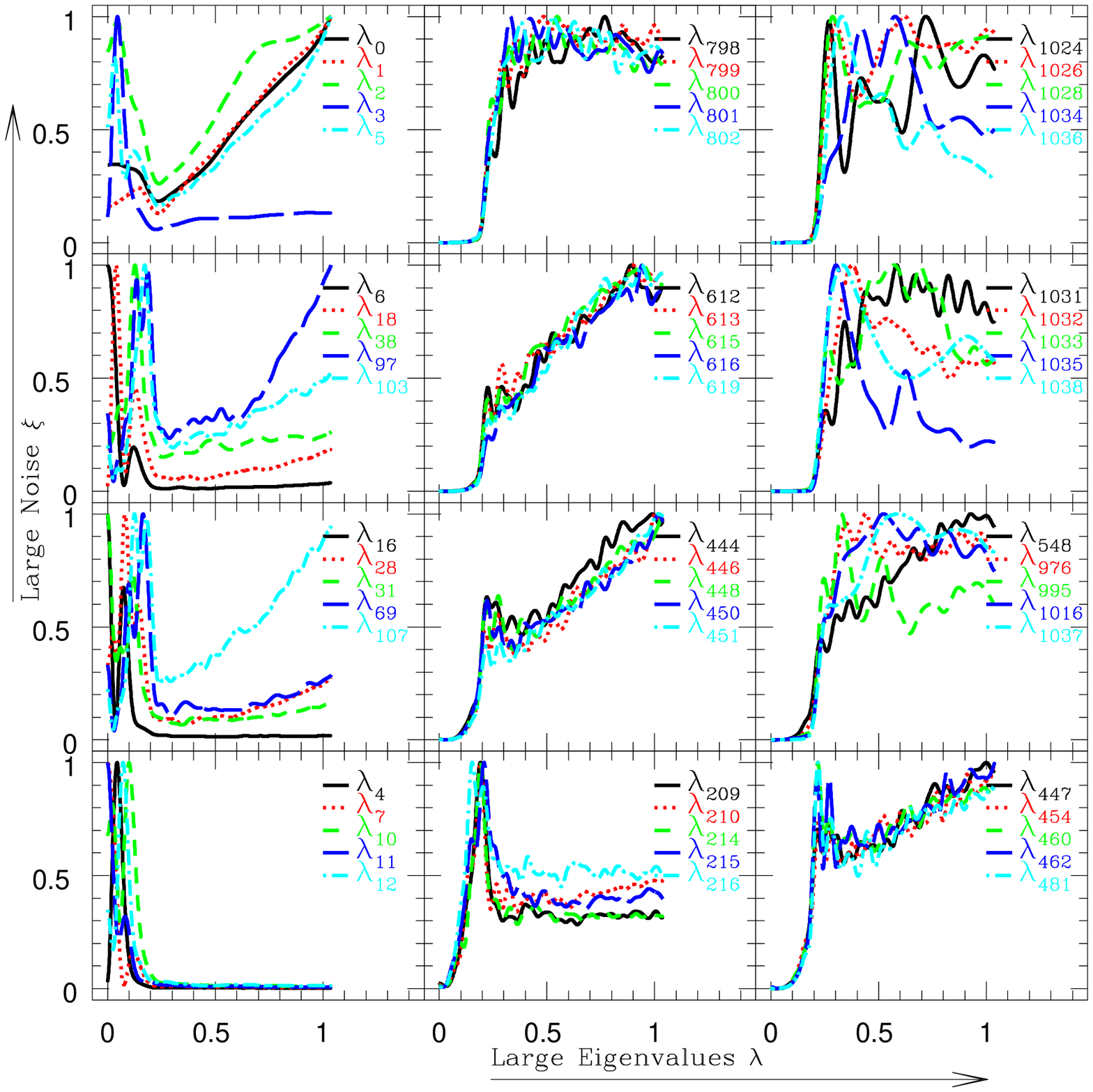,height=2.5in}\qquad\qquad\qquad
\psfig{figure=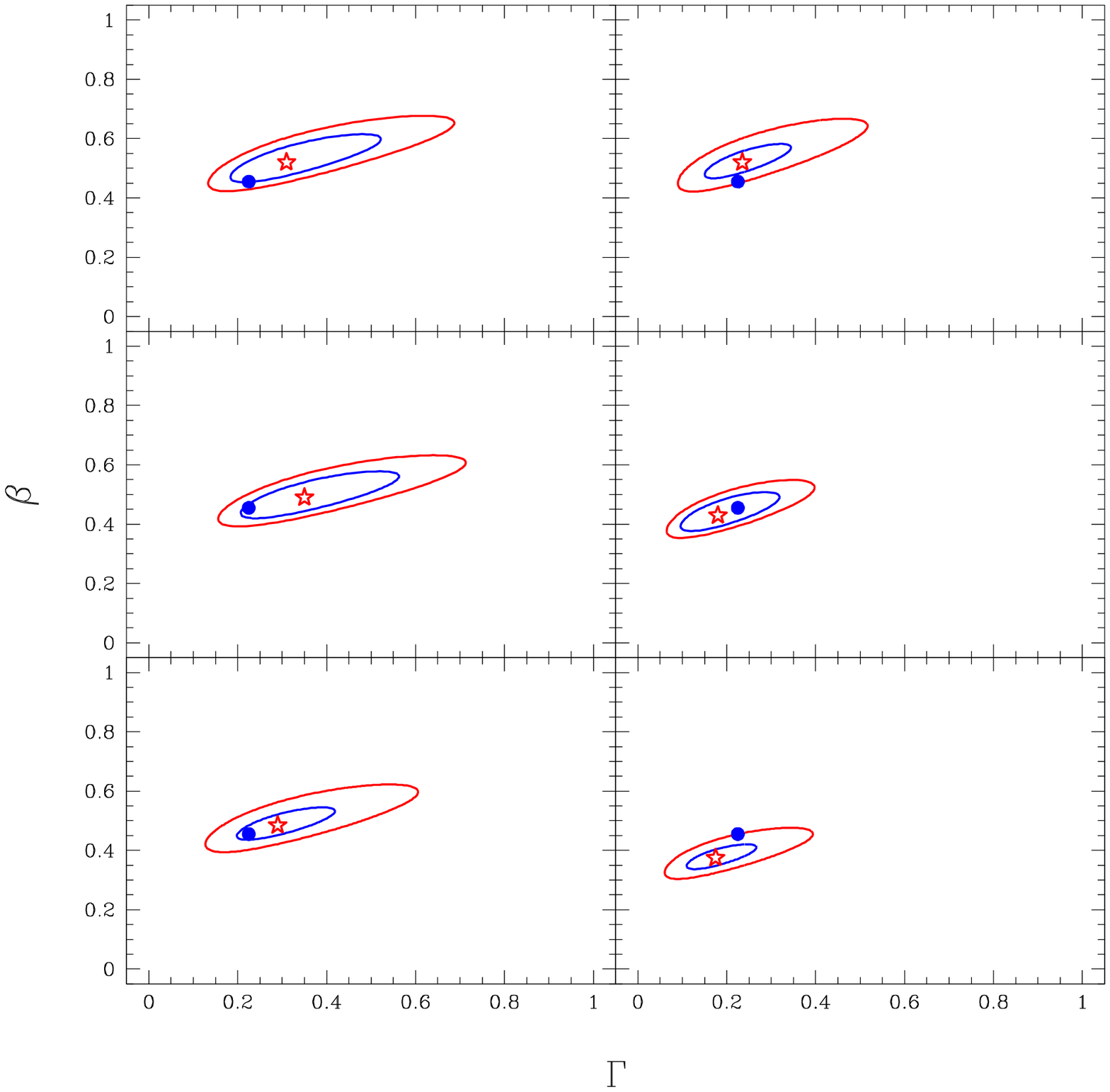,height=2.5in}
\end{center}
\caption{(left panel) The window
functions $W_n(k)$ in arbitrary units, from top to bottom
corresponding to noise in the ranges of $0.98<\xi$, $0.95<\xi<0.98$,
$0.9<\xi<0.95$ and $\xi<0.9$ respectively, and across from left low,
medium and high eigenvalues $\lambda$ respectively. We can clearly see
that the low eigenvalue low noise window functions (lower left panel)
probe large scale (small $k$), whereas higher noise, larger eigenvalue
window functions (up and to the right) correspond to smaller scales
probes. \hfill\break
Figure 4: (right panel) The maximum likelihood contours from six typical mock catalogs. The contours
are the 68\% and 94\% likelihood lines.  In most cases the
uncertainties in the estimated values of the parameters $\Gamma$ and
$\beta$ are of comparable sizes to the monte carlo errorbars presented
in figures~\ref{fig-1-2}
} 
\label{fig-7-9}
\end{figure} 

In Fig.~4 we show the contours that contain 68\% and
94\% of the total likelihood for six typical catalogs. The diamond shows
the maximum likelihood results, whereas the asterisk in each panel shows
the ``true'' values of the parameters. These contours allow us to
estimate the uncertainty in the maximum likelihood values obtained from
the analysis of a single catalog, as is the case when analyzing
observational data. From the figures it is clear that the uncertainties
obtained in this way are comparable to those we get from the
Monte--Carlo simulations. In general, when we try to test the
reliability of results from an observational data set, we apply our
formalism to mock catalogs extracted from N--body simulations as was
done here.  This compatibility between the uncertainties obtained in two
different ways gives us confidence that using the likelihood contours
will give us an accurate assessment of the uncertainties of our maximum
likelihood values when we apply our method to real catalogs.

%\section{Conclusions}
%\label{conclusions}

We have described the power and elegance of a new statistic that was
designed and formulated in order to address a crisis in the analysis of
proper distance cosmological surveys. We have shown that our formalism
mostly overcomes the problems with the traditional analysis of the
data. Whereas the full maximum likelihood analysis tends to systematically overestimate
the values of the parameters that describe the power distribution on
large scale, our mode analysis makes very accurate estimates of these
parameters.

As was shown in our recent publications~\cite{PI,PII}, the formalism is highly adaptive and versatile. It can be applied surveys with any geometry and density, and since it retains maximum information
should be particularly useful for sparse data such as that obtained in
cluster peculiar velocity surveys.  Overall, we consider this method to
be a significant improvement over previous methods used for the analysis
of peculiar velocity data.

\section*{Acknowledgments}

HAF and ALM wish to acknowledge support
from the National Science Foundation under grant number AST--0070702,
the University of Kansas General Research Fund and the National Center
for Supercomputing Applications for allocation of computer time. This
research has been partially supported by the Lady Davis and Schonbrunn Foundation at
the Hebrew University, Jerusalem, Israel and by the Institute of
Theoretical Physics at the Technion, Haifa, Israel.

\end{document}